\documentclass[%
 preprint,
 superscriptaddress,
 amsmath,amssymb,
 aps,
 showkeys,
 titlepage,
 endfloats*.   
]{revtex4-2}

\usepackage[utf8]{inputenc}
\usepackage[english]{babel}

\usepackage{graphicx}
\usepackage{dcolumn}
\usepackage{bm}

\usepackage[colorlinks = true,
            linkcolor = blue,
            urlcolor  = blue,
            citecolor = red,
            anchorcolor = blue]{hyperref}

\begin{document}


\title{Enhancing Magnetocaloric Material Discovery: A Machine Learning Approach Using an Autogenerated Database by Large Language Models} 

\author{Jiaoyue Yuan}
\thanks{These authors contributed equally.} \affiliation{Department of Mechanical Engineering, University of California, Santa Barbara, CA 93106, USA}
\affiliation{Department of Physics, University of California, Santa Barbara, CA 93106, USA}

\author{Runqing Yang}
\thanks{These authors contributed equally.} \affiliation{Department of Mechanical Engineering, University of California, Santa Barbara, CA 93106, USA}

\author{Lokanath Patra}
\thanks{These authors contributed equally.} \affiliation{Department of Mechanical Engineering, University of California, Santa Barbara, CA 93106, USA}

\author{Bolin Liao}
\email{bliao@ucsb.edu} \affiliation{Department of Mechanical Engineering, University of California, Santa Barbara, CA 93106, USA}

\date{\today}

\begin{abstract}
Magnetic cooling based on the magnetocaloric effect is a promising solid-state refrigeration technology for a wide range of applications in different temperature ranges. Previous studies have mostly focused on near room temperature (300\,K) and cryogenic temperature ($<$ 10\,K) ranges, while important applications such as hydrogen liquefaction call for efficient magnetic refrigerants for the intermediate temperature 10\,K to 100\,K. For efficient use in this range, new magnetocaloric materials with matching Curie temperatures need to be discovered, while conventional experimental approaches are typically time-consuming and expensive. Here, we report a computational material discovery pipeline based on a materials database containing more than 6000 entries auto-generated by extracting reported material properties from literature using a large language model. We then use this database to train a machine learning model that can efficiently predict magnetocaloric properties of materials based on their chemical composition. We further verify the magnetocaloric properties of predicted compounds using \textit{ab initio} atomistic spin dynamics simulations to close the loop for computational material discovery. Using this approach, we identify 11 new promising  magnetocaloric materials for the target temperature range. Our work demonstrates the potential of combining large language models, machine learning, and \textit{ab initio} simulations to efficiently discover new functional materials.
\end{abstract}

                            
\maketitle


\section{Introduction}
Solid-state refrigeration presents a groundbreaking avenue driven by the pressing need for more sustainable and efficient cooling technologies~\cite{pecharsky1999magnetocaloric,franco2012magnetocaloric, gutfleisch2011magnetic, balli2017advanced}.
Specifically, magnetic refrigeration shows promise in solid-state cooling applications across a wide temperature range. Magnetic refrigeration is based on the magnetocaloric effect (MCE), where the entropy of a magnetic material (the ``magnetic refrigerant'') is controlled by an external magnetic field~\cite{pecharsky1999magnetocaloric}. When an external magnetic field is applied, the microscopic magnetic moments are aligned with the field. As a result, the entropy associated with the magnetic moments in the material is reduced. In the process, either heat is released under an isothermal condition or the temperature of the magnetic refrigerant is increased under an adiabatic condition. When the magnetic field is removed, the effect is reversed. These processes can be combined to form a thermodynamic cooling cycle, whose cooling capacity is determined by the isothermal magnetocaloric entropy change ($\Delta S_M$) and the adiabatic temperature change ($\Delta T_{ad}$) during the magnetization and demagnetization processes. This cycle resembles a Carnot cycle on the $S$-$T$ diagram, promising close-to-ideal efficiency~\cite{lyubina2017magnetocaloric}. Compared with other cooling technologies, magnetic refrigeration has received renewed interest thanks to its solid-state nature, cryogen-free operation, reliability, compactness, and capability to reach close-to-Carnot efficiency at cryogenic temperatures~\cite{shirron2007cooling}. For these reasons, magnetic refrigeration has been widely applied for applications in space missions~\cite{shirron2012design}, scientific instrumentation~\cite{sato2016tiny}, and as an environmentally friendly alternative to vapor compression-based cooling~\cite{aprea2015magnetic}. However, most previous investigations of magnetic refrigeration have focused on either cryogenic temperatures below a few K or near room temperature, while many technologically important applications require efficient cooling in the temperature range between 10\,K and 100\,K, such as hydrogen and methane liquefaction~\cite{teyber2019performance} and cryogenic carbon capture~\cite{safdarnejad2015plant}.  

To achieve efficient magnetic refrigeration in this temperature range, one key requirement is to identify new high-performance magnetic refrigerants with their Curie temperature ($T_C$) in this range, since both $\Delta S_M$ and $\Delta T_{ad}$ are typically maximized near $T_C$. Therefore, $T_C$ signifies the ideal working temperature for an MCE material and stands as a crucial parameter for identifying new magnetic refrigerants. In addition, large magnetic moments and favorable exchange interactions~\cite{patra2023indirect} are fundamental prerequisites for a material to exhibit favorable MCE characteristics. In this context, rare-earth-based alloys are state-of-the-art candidates for magnetocaloric refrigeration in the intermediate temperature range. For example, heavy rare-earth-based MCE materials, such as HoB$_2$,~\cite{castro2020machine,terada2020relationship} ErCo$_2$,~\cite{wada2001magnetocaloric} DyAl$_2$,~\cite{von1998influence, von2008giant} and ErAl$_2$,~\cite{von1998influence} exhibit significant $\Delta S_M$ and $\Delta T_{ad}$ values in this temperature range due to their large magnetic moments. Consequently, they are actively researched for applications like hydrogen liquefaction because they have $T_C$ within the required range~\cite{acosta2022machine,numazawa2014magnetic,kim2013experimental}. Their typical isothermal entropy change $\Delta S_M$ is around 20-30 J\,kg$^{-1}$K$^{-1}$ and adiabatic temperature change $\Delta T_{ad}$ is around 10\,K for an applied magnetic field of 5\,T. Historically, these MCE materials were discovered mainly through an experimental synthesis and characterization process, which can be expensive, time-consuming, and often relies on trial and error. 




To make further progress, more efficient methods to discover and verify new high-performance MCE materials are needed for refrigeration applications in various temperature ranges. Along this line, high-throughput computational material searches were previously conducted~\cite{bocarsly2017simple,arapan2018high}. For example, Bocarsly et al. used the strength of spin-lattice coupling in a material, which can be effectively evaluated at the ground-state density functional theory (DFT) level, as a proxy to search for new magnetocaloric materials~\cite{bocarsly2017simple}. In addition, previous studies have used various machine learning-based methods to predict suitable candidates for magnetic refrigeration from the aspects of compositional and structural changes~\cite{zhang2020machine} and optimization to enhance key performance metrics while reducing material cost~\cite{kusne2014fly, silva2020predicting}. Suitable candidates such as HoB$_2$ and Co-doped Fe$_2$P-type compounds were identified for this purpose by data mining and machine learning ~\cite{castro2020machine, lai2022machine}. However, these pioneering studies have been limited in the size of accessible MCE datasets and have lacked an efficient way to verify the MCE performance of predicted materials to close the loop for material discovery. As a result, new MCE materials of practical value that have been identified through these approaches have been limited.  

In the realm of materials science, there is a vast and expanding collection of unstructured data in the form of journal articles, patents, and theses. This accumulation and rapid publication pace overwhelm researchers' ability to fully utilize the available information. Converting these data into structured materials databases would facilitate data-driven research aimed at discovering new materials, yet manual compilation is not only labor-intensive, but also susceptible to errors. Fortunately, advancements in natural language processing (NLP) have enabled the automated extraction of data from extensive document collections, paving the way for creating detailed materials property databases and discovering new materials~\cite{olivetti2020data,venugopal2021looking}. For example, in a previous study, Court \textit{et al.} have established an auto-generated database of 39,822 records containing magnetic materials and their associated Curie or N\'{e}el magnetic phase transition temperatures using NLP to extract relevant information from the literature~\cite{court2018auto}. In a subsequent study, they used an NLP toolkit to generate a material database from published literature containing 2910 entries with their MCE properties and then used the data to train a deep neural network to predict MCE properties of new materials~\cite{court2021inverse}. They identified six new promising MCE materials with this innovative approach. However, the validation of their results still relied on experiments that can be the bottleneck of the pipeline. 

Two recent advancements have provided new opportunities to further improve the pipeline proposed by Court \textit{et al.}~\cite{court2021inverse} to efficiently and accurately identify new MCE materials. One is the rapid rise of large language models (LLMs), particularly those in the Generative Pre-trained Transformer (GPT) series~\cite{min2023recent}, which have demonstrated exceptional capabilities in generating coherent and contextually appropriate text and have also introduced advanced features such as summarizing articles, generating informative charts, and mapping relationships between words. The emergence of LLMs enables efficient extraction of material properties from a large amount of published literature~\cite{jablonka202314,polak2024extracting}. The other development is computational methods based on atomistic spin dynamics (ASD) simulations with exchange interaction coefficients computed from first principles~\cite{antropov1995ab,skubic2008method,patra2023indirect}.
These methods allow an accurate evaluation of MCE properties based solely on the crystal structure and chemical composition of materials and can close the loop in high-throughput and machine learning-based computational material searches.

\begin{figure}[!htb]
\includegraphics[width=1\textwidth]{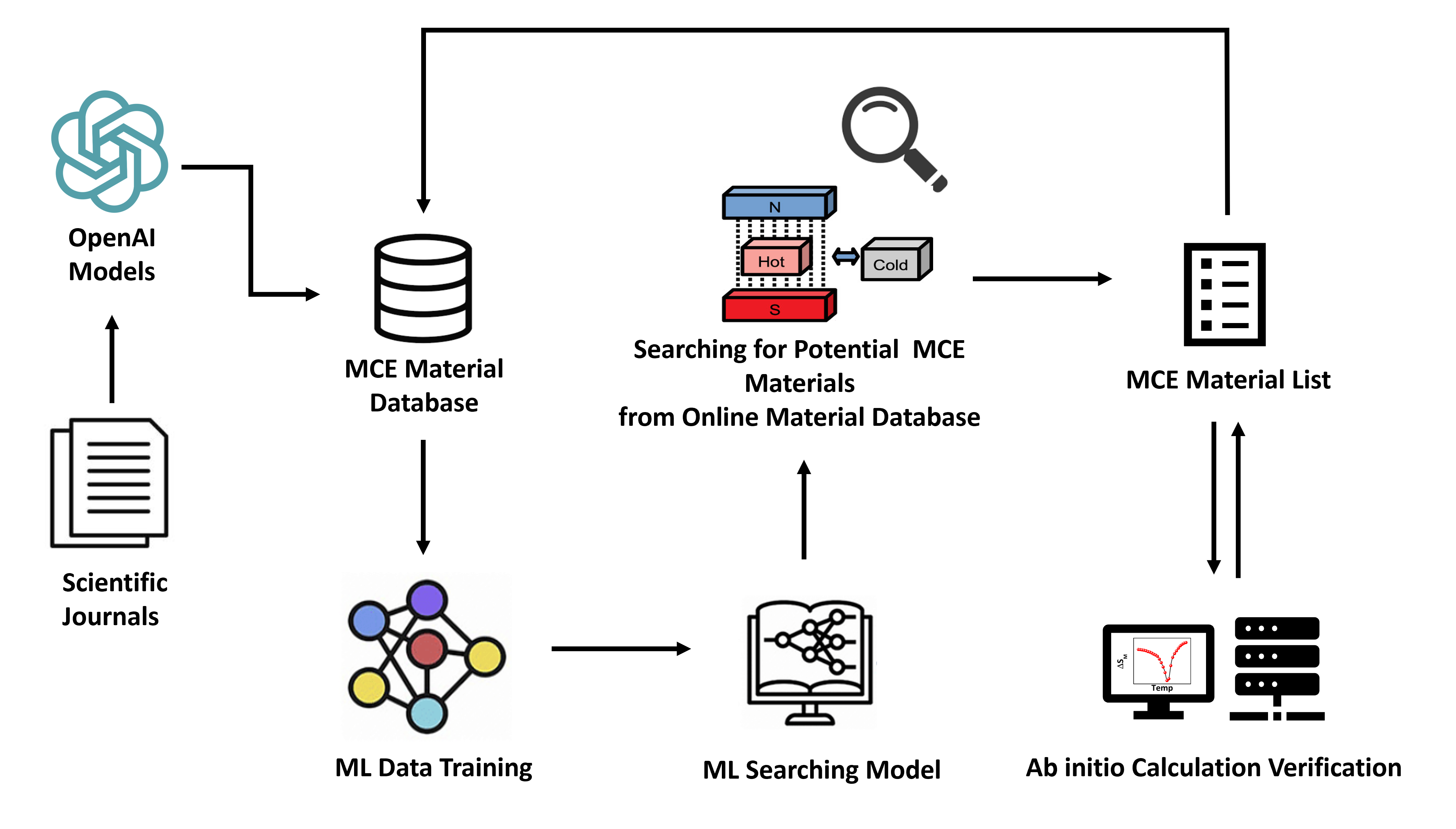}
\caption{\textbf{Flowchart of the magnetocaloric material discovery pipeline}. Existing literature is analyzed by a large language model to generate a database containing the MCE properties of a large number of materials. The database is then curated and refined to serve as the training dataset for a machine learning model. The machine learning model is then used to screen materials in the Materials Project database to identify new promising MCE materials, whose properties are verified by ab initio DFT and ASD simulations. The simulation results are added back to the material database to form a closed loop to refine the process. }
\label{fig:fig1}
\end{figure}

\begin{figure}[!htb]
\includegraphics[width=1\textwidth]{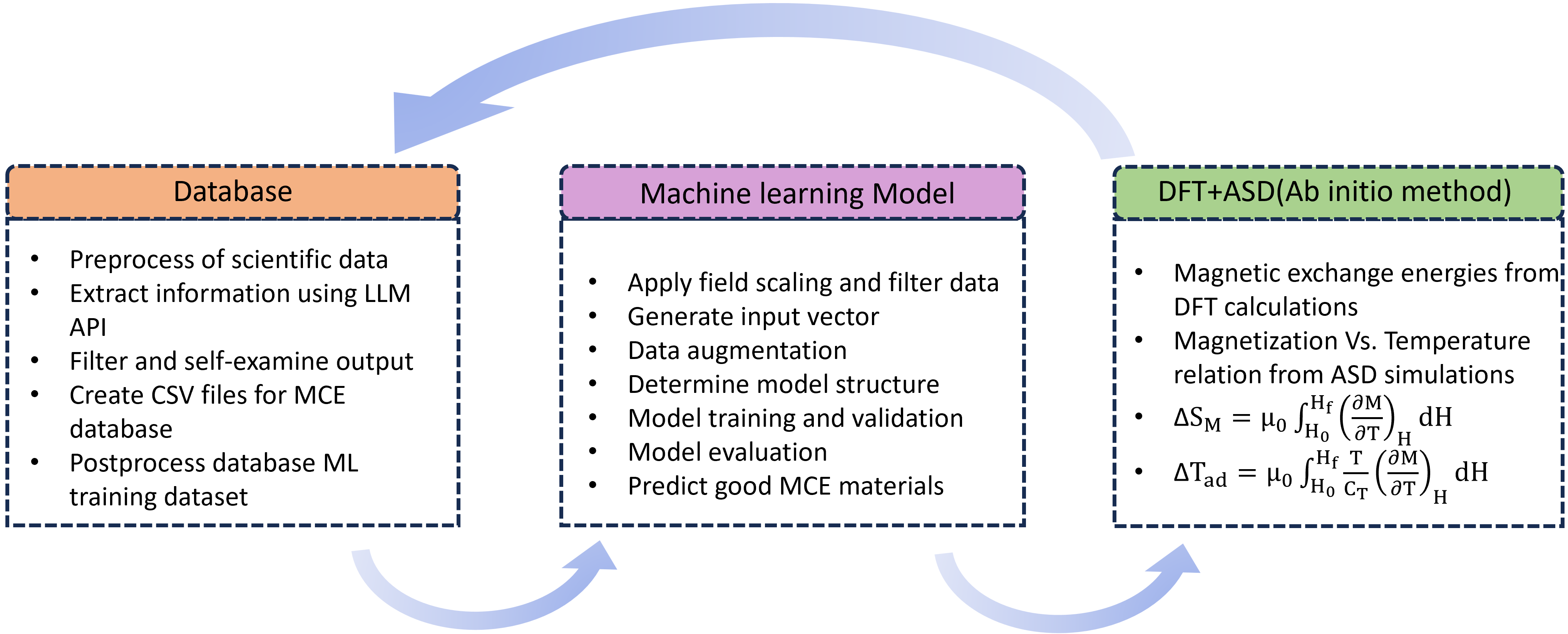}
\caption{\textbf{Detailed information on the key steps of the magnetocaloric material discovery pipeline.} }
\label{fig:fig2}
\end{figure}

In this work, we exploit GPT's remarkable abilities to summarize texts to complete an auto-generated MCE material database with 6615 entries based on openly accessible abstracts of published literature. After careful curation and validation, we used high-quality data from the database to train a neural network that can predict the isothermal entropy change of ferromagnetic materials with high fidelity only based on their chemical composition. With this neural network model, we efficiently screened binary and ternary magnetic compounds in the large online database, the Materials Project~\cite{jain2013commentary}, and identified a list of promising MCE materials that have not been studied before. Finally, we used DFT-based first-principles simulation and ASD to evaluate the MCE properties of the top-ranked candidates predicted by the neural network and identified 11 new promising MCE materials for the less explored intermediate temperature range (10\,K to 100\,K). Importantly, the calculated results can be added back to the MCE database which can further improve the performance of the neural network in the fashion of a closed feedback loop. A schematic of our workflow is shown in Fig.~\ref{fig:fig1} with more detailed information about each key step summarized in Fig.~\ref{fig:fig2}. Our work demonstrates the vast potential of using LLMs to extract structured information from published literature and utilize the extracted data to drive new discoveries of functional materials. 

\section{Methods}

\subsection{Database Creation using ChatGPT}
To further the discovery of new MCE materials through machine learning, it is crucial to establish a comprehensive and high-quality database containing the MCE properties of materials that can be then used as the training dataset for a machine learning model. 
Fortunately, the rapid advancement of LLMs, particularly with the introduction of ChatGPT by OpenAI in 2022, offers a promising solution to automatically generate a database based on the large body of published literature \cite{dagdelen2024infoextract}.
The process for creating our MCE database using ChatGPT is detailed below:
\\
\textbf{Data Collection:} We began by selecting the first 8,000 most relevant abstracts from the Web of Science from more than 350 journals, ranking all scientific papers by their relevance to the MCE properties of materials.
The time span of the publication date extends from 1993 to 2023.
These abstracts were then converted into CSV files containing both the text of the abstract and publication details. 
To streamline this process, we focused on papers that provided useful data, such as those containing one or more chemical compounds and their corresponding MCE properties expressed in numerical values. To reduce the number of records to a more tractable quantity for further analysis, we developed a filter pipeline that employs prompts to determine the potential usefulness of data within an abstract. 
This filtering process is both accurate and efficient, as commercialized LLMs like GPT can readily identify the presence of chemical compounds or numerical values in the text. Abstracts that only offer general descriptions or values without specific materials were excluded.
\\
\textbf{Prompt Engineering and Self-Examination:} Utilizing the OpenAI's LLM API, we transformed abstracts into tables, highlighting material compositions and their properties. This involved using a multi-shot prompt strategy~\cite{logan2021pompt,roemer2024pompt}, which increases the accuracy of data extraction by providing the LLM with multiple examples. 
The tables generated were generally accurate. GPT models, like other LLM models, exhibit a phenomenon called hallucination, wherein they can generate low-likelihood outputs that deviate from the expected output based on the input context and the true data distribution~\cite{math11102320}.
To address this potential issue, we applied a self-examination pipeline that aims to reduce the likelihood of errors due to hallucinations, such as the inclusion of unmentioned materials, overlooking properties, or confusing property names.
The self-examination process involves an additional prompt to verify the consistency between the generated table and the original abstract, specifically to counteract the LLM's tendency to make educated guesses by removing numerical values not provided in the abstract. These steps were conducted separately, guiding the LLM to focus on task reasoning rather than relying on its vast knowledge base.
\\
\textbf{Filtering after Natural Language Processing}: After extracting relevant material formulations and MCE properties from the literature using GPT, we established a database with MCE properties for 6615 materials. Starting from this collection, we refined the dataset by filtering out organic materials and alloyed materials to limit our material discovery process to intrinsic inorganic materials as the starting step, which leads to a reduced database with 1426 entries. In principle, however, the material search can be expanded later to include more complex materials in our future work.
\\
\textbf{Database Creation and Mapping to the Materials Project:} Our objective was to create a one-to-one match between the materials listed in our filtered database and those cataloged in the esteemed Materials Project (MP) database~\cite{jain2013commentary}. 
This approach was chosen because the MP database provides a wealth of physical properties data (notably, without the MCE properties), which, when combined with our MCE data, facilitates their use in machine learning for materials discovery. 
Ultimately, we successfully matched 752 materials in our database with their corresponding Material Project IDs. 
To enhance accessibility and utility, we developed an HTML version of the database~\footnote{\url{https://github.com/RunqingYang1996/MCEdatabase}}. 
This allows users to efficiently search for reported MCE materials using various criteria, such as chemical elements, or to organize materials by their name, MCE properties, or Curie temperatures.
\\
\textbf{Evaluation:} 
We manually reviewed 200 randomly chosen samples from each test set to assess the database's accuracy. The database was estimated to have an accuracy of around 90$\%$. A record is considered valid only if the name of one material matches the correct magnetocaloric entropy change ($\Delta S_M$) and $T_C$ values cited in the input scientific text. Information about our database is summarized in Table \ref{tab:database}.

\begin{table}[htb]
\caption{\label{tab:database}
Information about the created and filtered MCE databases.
}
\begin{ruledtabular}\scriptsize
\begin{tabular}{c|c|c} 
Number of Records & Database Information  & Estimated Precision \\ 
\hline \hline 
6615 & Material name, $\Delta S_M$ and/or $T_C$. Organic and alloyed materials are included. & 85\% \\ 
\hline
1426 & Material name, $\Delta S_M$ and/or $T_C$. Only intrinsic inorganic materials are included. & 90\% \\ 
\hline
752 & Material name, $\Delta S_M$ and $T_C$, and Materials Project ID. & 90\% \\ 
\hline
752 & Human-corrected version of the 752-record database for machine learning training  & 99\% \\
\end{tabular}
\end{ruledtabular}
\end{table}

\subsection{Machine Learning Model Construction and Training}

Our next step is to train a machining learning model using the database constructed by the LLM to predict MCE properties of materials based on their chemical composition. One challenge arises that the reported $\Delta S_M$ values were measured across varying magnetic field strengths in different papers, rendering them incomparable.
To address this, we implemented a linear scaling based on a commonly used field strength of 5\,T to approximate the field dependence of $\Delta S_M$ as established in measurements~\cite{hcini2019structural}.
Another issue is that the dataset contains a notable number of duplicates due to measurements by different groups or at different field strengths. We kept the maximum value among the duplicates after scaling to make $\Delta S_M$ values consistent for training. To leverage data efficiency and model complexity, we limited the training data to include only materials with less than or equal to 8 atoms in the unit cell, resulting in 229 unique materials in the training set. We adopted a feedforward neural network with 3 hidden layers, each with 16 neurons. The optimal network structure was determined via a grid search of different numbers of layers and neurons per layer. Details of the process can be found in Supplementary Information (SI) Sec.~I.B~\cite{SI}. The input vector contains 4 features with a total vector size of 11: 8 integers representing the atoms in the unit cell, and 3 floats for volume, density, and total magnetization, respectively, which can be extracted from the Materials Project database. In general, neural networks require larger amounts of high-quality training data, enabling them to learn the underlying patterns and relationships. In our case, a major challenge is that we achieved high training data quality at the expense of the total number of records. In this study, we tackled this by employing data augmentation. Since the order of atomic numbers in the vector representation can be arbitrary, permutations of atomic numbers essentially represent the same material. Hence, we shuffled the elements of the atomic number vector to create multiple representations and augment the size of the training set.

\begin{figure}[!htb]
\includegraphics[scale=0.5]{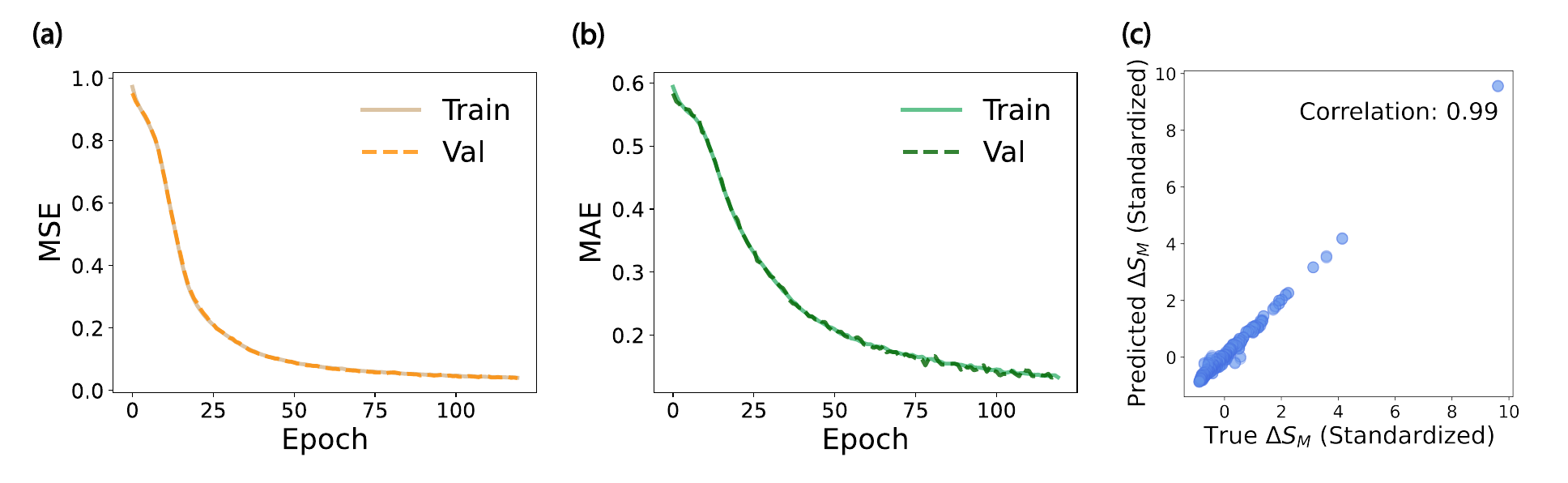}
\caption{\textbf{Training and the performance metric of the machine learning model.} Training curves of (a) the mean-squared error (MSE) and (b) the mean-absolute error (MAE) of the machine learning model. Here the metrics for the training set and the validation set are plotted in solid and dashed lines, respectively. (c) Correlation plot of the predicted standardized $\Delta S_M$ by the trained machine learning model versus the true standardized $\Delta S_M$ in the database.} 
\label{fig:fig3}
\end{figure}

In the model, we generated 100 ``representations'' of each unique material as training data.
Test runs that validated the proof of concept are included in SI Figs. S2 and S3 ~\cite{SI}. The training process aims to minimize the mean squared error (MSE), which is more susceptible to outliers because it penalizes larger errors more significantly.
In this context, MSE is suitable for the purpose of identifying candidates with large $\Delta S_M$ values.
Standardization is applied to the dataset, which is a common preprocessing step in machine learning to help with algorithm convergence.
It ensures that each column of input features and output has zero mean and unit variance.
The target of the model, standardized $\Delta S_M$ values, is directly used to rank the candidate materials.
The predictions can be converted back to the original scale by an inverse transformation.
Fig.~\ref{fig:fig3} presents the training curves and an evaluation of the adopted model. The first two panels show the loss function (MSE) and the mean absolute error (MAE) of the training process, indicating the model's accuracy and convergence. Notably, the consistently low values of both MSE and MAE across the panels affirm the robustness and accuracy of the adopted model. The third panel illustrates the correlation between predicted $\Delta S_M$ and true $\Delta S_M$ values of the dataset, showing the model's predictive capabilities.  For each data point, the atomic number vector of the input representation was shuffled 10 times, and the average of the 10 predictions was plotted.  The high correlation coefficient of 0.99 underscores a strong alignment between the predicted values by the neural network and the ground truth. This impressive correlation emphasizes the model's reliability in capturing patterns within the data and reinforces its suitability for accurate predictions. 

We executed the model 10 times to assess the consistency of rankings and produce the final ranking. 
The ranking from each run is recorded in a sorted list for each material, and this sorted list is used for the final ranking based on its length and each ranking. We evaluated the model performance by the ``recall'' of the ranked list~\cite{sammut2011encyclopedia}. Recall is a performance metric in machine learning that measures the true positive rate of a model’s prediction. In other words, it measures the proportion of actual positive cases that the model correctly identifies as positive.  For example, ``recall@10'' measures that, among the top 10 candidates predicted by the model, the percentage that accurately aligns with the top 10 candidates according to the ground truth. Table \ref{tab:Model} shows the recall when considering the top $K$ candidates. ``Max ranking'' indicates the number of top candidates considered in producing the sorted list, which can only be smaller or equal to $K$. By examining the recalls at various $K$ values, it becomes evident that our model demonstrates consistent performance across different ranking strategies.

\begin{table}[htb]
\caption{\label{tab:recall}
\label{tab:Model}
Recall of top K materials ranked by $\Delta S_M$ value.
}
\begin{ruledtabular}
\begin{tabular}{c|cccc}
Max Ranking / K & 10 &  25 &  50 \\
\hline
\hline
10 & 1.0 & 1.0 & 1.0 
\\
\hline
25 & - & 0.96 & 0.96 
\\
\hline
50 & - & - & 0.96
\\
\end{tabular}
\end{ruledtabular}
\end{table}

\subsection{DFT and ASD Simulations}
Once new promising MCE materials are predicted by the machine learning model, we use DFT and ASD simulations to directly compute their MCE properties ($\Delta S_M$ and $\Delta T_{ad}$) and verify the performance of the machine learning model. For each magnetic material, the magnetic exchange energies ($J_{ij}$) were estimated with DFT calculations using the Vienna ab initio simulation package (VASP) based on the projected augmented wave pseudopotentials~\cite{kresse1996efficient}. The Perdew-Burke-Ernzerhof form of generalized gradient approximation (PBE-GGA)~\cite{blochl1994projector} was used for structural optimization. All of our calculations utilized a plane-wave cut-off energy of 600 eV. The energy and force convergence criteria were set to be 1 $\times$ 10$^{-5}$ and 0.01 eV/\AA, respectively. Monkhorst-Pack~\cite{monkhorst1976special} \textbf{k}-point grid of (20 $\times$ 20 $\times$ 20) was employed for cubic GdMg, and similar density of \textbf{k}-point grids were utilized to sample the Brillouin zone for all other compounds. The robust correlation among the localized `$f$' electrons was systematically addressed via GGA+U calculations. Specifically, the parameters $U$ and $J$ were set to 7 eV and 1 eV, respectively, for all rare-earth elements in our calculations, as these values are commonly used in the literature~\cite{duan2007electronic,wang2019r3c}. Three nearest-neighbor $J_{ij}$ values were estimated in each case and subsequently utilized for ASD simulations in further steps.

\begin{figure}
    \centering
    \includegraphics[width=1\linewidth]{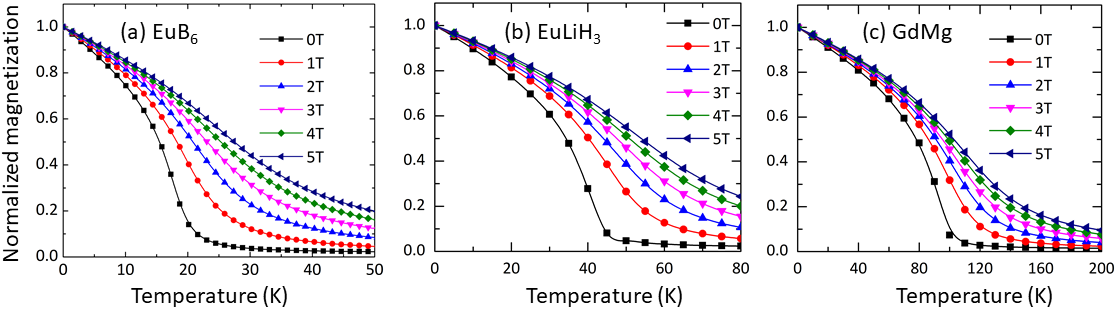}
    \caption{\textbf{Simulated temperature and field-dependent magnetization in three representative compounds predicted by the machine learning model.}}
    \label{fig:MCE-Tc}
\end{figure}

The parameterized atomistic spin model was employed for ASD simulations using VAMPIRE simulation software~\cite{evans2014atomistic}. To determine the $T_C$, the Monte Carlo method was utilized. This involved executing 10,000 steps at each temperature, allowing the system of $30 \times 30 \times 30$ unit cells with in-plane periodic boundary conditions to reach equilibrium. Subsequently, a statistical average was obtained over 10,000 steps to extract the mean magnetization. Temperature-dependent magnetization curves were calculated employing the spin dynamics approach and the Heun integration scheme~\cite{garcia1998langevin}. Additionally, the demagnetization field induced by the atomistic spins themselves was accounted for. The external magnetic field was systematically increased to 5\,T with an incremental step of 1\,T. The ASD simulation was conducted to generate the magnetization versus temperature ($M$ vs. $T$) curves. Figure~\ref{fig:MCE-Tc} represents the $M$ vs. $T$ curves for three representative materials i.e. EuB$_6$, EuLiH$_3$, and GdMg. From these temperature-dependent magnetization curves, $T_C$ values were estimated by fitting the curves with $M(T) = \left(1 - \frac{T}{T_C}\right)^\beta$, where $\beta$ is the critical exponent. The $M$ vs. $T$ curves for fields ranging from 0 to 5\,T were calculated with steps of 1\,T to evaluate $\Delta S_M$  using the Maxwell relation~\cite{pecharsky1999magnetocaloric}:
 \begin{equation}
     \Delta S_M = \mu_0 \int_{H_i}^{H_f} \left( \frac{\partial M }{\partial T} \right)_H dH,
 \label{eqn-dsm}
 \end{equation}
 where $H_i$ and $H_f$ represent the initial and final magnetic fields, respectively, and $\mu_0$ denotes the Bohr magneton. For further evaluation of the MCE performance of the predicted compounds, the total  specific heat ($C_T$) was estimated as a combination of phononic ($C_p$), magnonic ($C_m$), and electronic ($C_e$) contributions, i.e.,
 \begin{equation}
     C_T = C_m + C_p + C_e.
 \end{equation}

 First, we begin with the magnetic energy ($U$) of the magnetic compound to calculate the $C_m$ i.e.
 \begin{equation}
     U = -\sum_{ij}\lambda_{ij}\Vec{J_i} \cdot \Vec{J_j}-\sum_{i}g\mu_B\Vec{J_i}\cdot h^{ext},
 \end{equation}
 where the first term represents the spin-spin exchange interaction, where $\lambda_{ij}$ is the exchange interaction parameter and $J_i$ is the total angular momentum of the magnetic ions. The second term represents the Zeeman interaction of the total angular momentum with an external magnetic field ($h^{ext}$). For a given temperature, the mean value of the energy $\langle U \rangle$ is calculated by:
 \begin{equation}
     \langle U \rangle = \frac{1}{(N_C-N_0)}\sum_{i>N_0}^{N_C}U_i,
     \label{eqn-U-mean}
 \end{equation}
where $N_C$ represents the total number of Monte Carlo cycles and $N_0$ is the number of Monte Carlo cycles used for thermalization. The mean square energy $\langle U^2 \rangle$ is obtained by a relation similar to Equation~\ref{eqn-U-mean}. Now, the magnetic part of the heat capacity can be calculated as
\begin{equation}
    C_m(T,h^{ext})=\frac{\langle U^2 \rangle - \langle U \rangle^2}{k_BT^2},
\end{equation}
where $k_B$ is the Boltzmann constant. In this study, the Monte-Carlo simulations were performed using the VAMPIRE simulation software~\cite{evans2014atomistic}, which produces the magnetic heat capacity as one of the outputs. 

Following the determination of phonon frequencies through the frozen phonon method,  constant volume phonon heat capacity ($C_V$) can be calculated using the following equation as implemented in the Phonopy code~\cite{phonopy-phono3py-JPCM,phonopy-phono3py-JPSJ}: 

\begin{equation}
    C_V = \sum_{qj} C_{qj} = \sum_{qj} k_B \left(\frac{\hbar \omega_{qj}}{k_BT}\right)^2 \frac{\exp (\hbar \omega_{qj}/k_BT)}{[\exp (\hbar \omega_{qj}/k_BT)-1]^2},
\end{equation}
where $q$ is the wave vector, $j$ is the band index, $\omega_{qj}$ is the phonon frequency, and $\hbar$ is the reduced Planck's constant. We used the calculated $C_V$ as an approximation for the constant-pressure heat capacity $C_p$, since in the low-temperature regime of solids, $C_V \approx C_p$~\cite{wang2012magnetic}.

Lastly, $C_e$ is given by:
\begin{equation}
    C_e = \frac{\pi^2k_B^2D(E_F)}{3}T = \gamma_eT,
\end{equation}
where $\frac{\pi^2k_B^2D(E_F)}{3}$ represents the electronic heat-capacity coefficient, and $D(E_F)$ represents the electronic density of states at the Fermi energy $E_F$. Next, the $\Delta T_{ad}$ was estimated using the calculated $C_T$ by employing the following equation:
 \begin{equation}
     \Delta T_{ad} = \mu_0 \int_{H_i}^{H_f} \frac{T}{C_T}\left( \frac{\partial M}{\partial T} \right)_H dH.
 \label{eqn-Tad}
 \end{equation}

\section{Results and Discussions}
\subsection{Database Generation and Machine Learning Model}
We successfully created a database with high accuracy using the LLM API, demonstrating the efficacy of our pipeline. The potential for hallucination (i.e., incorrect $\Delta S_M$ or $T_C$ values) was avoided through the self-examination process. The final database, containing 752 entries, is machine learning-friendly due to the clarity of material names and the ease of finding additional physical properties by cross-referencing the Materials Project database. However, there are still key issues that limit the precision of the database. Firstly, the text format cannot always accurately identify mathematical or physical symbols, leading to the loss of key information (e.g. numerical value or unit). Secondly, the LLM API sometimes fails to make correct judgments when the text mentions both Curie temperature and N\'eel temperature. As a result, some of the N\'eel temperature records within the database have been incorrectly identified as Curie temperatures, which is an intrinsic issue with the LLM. Compared to this, the LLM has high accuracy for $\Delta S_M$ values for each record. We used a human-corrected version of the database for higher accuracy training of the machine learning model. 

One strength of our model is that we added additional physical information to our input vectors obtained from the Materials Project database during the training process. Physical information is added by containing total magnetization, volume, and density information from the Materials Project and the spatial partition of the element distribution. The results showed robust performance-based MSE and MAE. We also did a recall evaluation of the model. In industrial applications, recall evaluation is frequently used where the emphasis is on identifying and capturing as many true positive cases as possible, even at the expense of some false positives. By examining the recalls at various $K$ values, it becomes evident that our model can consistently rank materials based on their $\Delta S_M$ values. Additionally, our model stands out for its remarkable efficiency.  Unlike many other studies that rely on complex model architectures or require extensive datasets with hundreds of features, this model achieves impressive results with a surprisingly simple structure. It utilizes a neural network with only 3 layers and 16 neurons and leverages a basic input vector of just 4 essential features with a total length of 11. This combination of simplicity and effectiveness makes this model a compelling choice for faster iteration and prediction.

The frequency distribution of $\Delta S_M$ and $T_C$ values of the training dataset are shown in SI Fig.~S1~\cite{SI}. It is noteworthy that 70.0\% of materials in the training dataset exhibit a $T_C$ below 100\,K. This prevalent feature is crucial for the model's capacity to identify additional candidates with a $T_C$ within the desired range (10\,K to 100\,K). The frequency distribution of rare-earth elements in our training dataset in SI Fig. 1c ~\cite{SI} shows that 81.6\% of reported MCE candidates in the dataset contain heavy rare-earth elements. This prior knowledge and observation motivated us to generate a screening list from the Materials Project by requiring the presence of heavy rare-earth elements: Gd, Tb, Dy, Ho, Er, Tm, and Eu. Using this criterion, we found 1124 stable materials containing 2 to 8 atoms in the unit cell. We separated them into three lists based on the number of elements: 370 candidates contain two elements, 711 candidates contain three elements, and 43 candidates contain at least four elements. The machine learning model provides one ranked list for each group. We picked the top 50 materials predicted by the machine learning model for further studies using DFT and ASD.

\subsection{DFT and ASD Calculations}
The primary objective of machine learning is material prediction; therefore, it is necessary to assess the quality of the predictive model. Hence, we validated the MCE response of the predicted compounds through DFT and ASD simulations. Initially, the predicted structures were optimized within the spin-polarized PBE-DFT level of theory. The total energies of magnetic structures with various combinations of in-plane and out-of-plane magnetic orderings were compared to determine the magnetic ground state. Materials exhibiting ferromagnetic ground states were further analyzed for their MCE properties, as they typically demonstrate direct MCE with $\Delta S_M < 0$ and $\Delta T_{ad} > 0$,~\cite{alho2022mean,von2009understanding} indicating that the sample heats up when an external magnetic field is applied adiabatically. Materials with a direct MCE effect typically exhibit higher magnitudes of magnetic entropy and adiabatic temperature changes compared to the inverse MCE observed in antiferromagnetic or ferrimagnetic materials~\cite{valiev2007entropy}.

\begin{figure}[!t]
\includegraphics[width=1\textwidth]{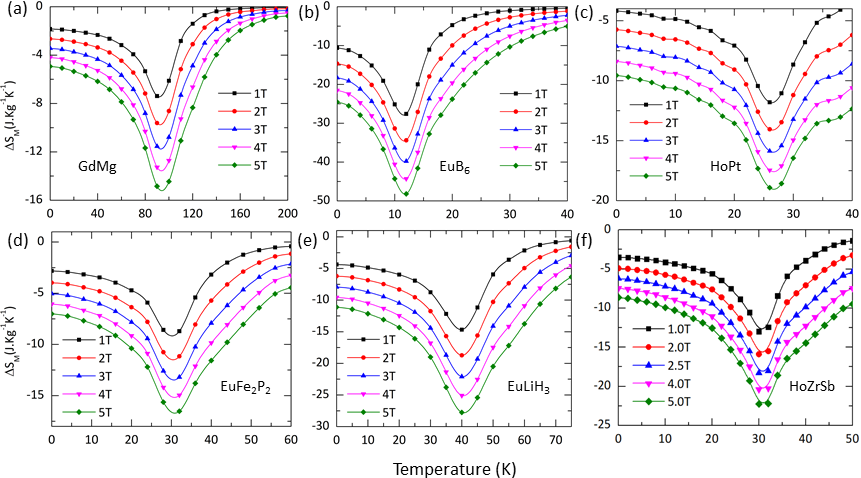}
\caption{\textbf{Simulated temperature and field-dependent magnetocaloric entropy change in six representative compounds predicted by the machine learning model.} } 
\label{fig:ASD}
\end{figure}

After running a magnetic ground state simulation based on DFT of the top 50 candidates predicted by our machine learning model, we identified 11 ferromagnetic compounds with their $T_C$ in the range 10\,K to 100\,K that have not been studied before for their MCE properties. We then conducted detailed DFT and ASD simulation of their MCE properties. The simulated temperature and field-dependent magnetization and the $\Delta S_M$ values of a few representative compounds are shown in Fig.~\ref{fig:MCE-Tc} and Fig.~\ref{fig:ASD}. The crystal structure symmetry, $T_C$, $\Delta S_M$, and $\Delta T_{ad}$ values evaluated using DFT and ASD for the machine learning-predicted compounds, are provided in Table~\ref{tab:ASD}. The estimated $T_C$ values closely match the reported measured values, considering that our simulations do not account for other effects such as anisotropy energy and crystal effective field effects.

Our computational predictions reveal a diverse array of magnetocaloric materials characterized by their $T_C$, ranging from the cryogenic temperatures of liquid helium (4.2\,K) to near the boiling point of liquid nitrogen (77\,K) (Table~\ref{tab:ASD}). These materials can be systematically classified into three distinct subgroups: lower (10 - 30\,K), medium (30 - 50\,K), and higher (50 - 90\,K) $T_C$ materials. Notably, the materials in the lower $T_C$ subgroup demonstrate exceptional potential for applications such as hydrogen liquefaction (typically around 20\,K) and low-temperature magnetic cooling. A notable candidate among these lower $T_C$ materials is EuB$_6$, which exhibits a theoretically predicted $\Delta S_M$ of approximately 48 J\,kg$^{-1}$K$^{-1}$, surpassing, to the best of our current knowledge, all other known materials in terms of mass-based performance within this temperature range~\cite{castro2020machine}. However, when considering the isothermal entropy change per volume, the $\Delta S_M$ of EuB$_6$ reduces to approximately 0.24 J\,cm$^{-3}$K$^{-1}$, a value comparatively lower than that of HoB$_2$, which exhibits a $\Delta S_M$ of approximately 0.36 J\,cm$^{-3}$K$^{-1}$~\cite{castro2020machine}. This disparity arises primarily due to the lower density of EuB$_6$ (5.01 g\,cm$^{-3}$) compared to HoB$_2$ (8.68 g\,cm$^{-3}$). Additionally, the estimated $\Delta T_{ad}$ for EuB$_6$ is approximately 8.1\,K, a figure comparable to other materials with similar transition temperatures~\cite{castro2020machine}. In the lower $T_C$ subgroup, HoPt exhibits the highest $\Delta S_M$ value in volumetric units, reaching 0.25 J\,cm$^{-3}$K$^{-1}$.

\begin{table}[tp]
\centering
\caption{The crystal structure, estimated (reported) $T_C$, and MCE properties for a field change of 5\,T of compounds predicted by the machine learning model.}
\label{tab:ASD}
\begin{ruledtabular}
\begin{tabular}{llccccc}
Compound & Structure  & $T_{C}$ (K)  & $\Delta S_M$ (J\,kg$^{-1}$K$^{-1}$) & $\Delta S_M$ (J\,cm$^{-3}$K$^{-1}$) & $\Delta T_{ad}$ (K)\\
\hline
EuB$_6$ & Cubic (Pm$\bar{3}$m)  & 12  (16)$^a$ & 48 & 0.24 & 8.1\\
TbCoC & Tetragonal (P4/mmm)  & 16 (22)$^b$ & 20 & 0.17 & 5.9\\
ErPt & Orthorhombic (Pnma)  & 24  (16)$^c$ & 14 & 0.20 & 4.5\\
HoPt & Orthorhombic (Pnma)  & 26  (16)$^c$ & 18 & 0.25 & 5.3\\
HoZrSb & Tetragonal (I4/mmm)  & 28 (22)$^d$ & 21 & 0.18 & 6.2\\
EuFe$_2$P$_2$ & Tetragonal (I4/mmm)  & 32 (30)$^e$ & 17 & 0.12 & 7.2\\
EuLiH$_3$ & Cubic (Pm$\bar{3}$m)  & 40  (38)$^f$ & 28 & 0.15 & 6.7\\
GdZrSb & Tetragonal (I4/mmm)  & 45 (37)$^d$ & 14 & 0.11 & 4.6\\
DyZrSb & Tetragonal (I4/mmm)  & 52 (45)$^d$ & 19 & 0.16 & 5.4\\
TbZrSb & Tetragonal (I4/mmm)  & 70 (61)$^d$  & 16 & 0.13 & 5.2\\
GdMg & Cubic (Pm$\bar{3}$m)  & 90  (120)$^g$ & 15 & 0.08 & 4.9\\
\end{tabular}
\end{ruledtabular}
\footnotesize{ $^a$\cite{sullow1998structure}, $^b$\cite{danebrock1995magnetic}, $^c$\cite{castets1980magnetic}, $^d$\cite{welter2003magnetic},  $^e$\cite{ryan2011magnetic}, $^f$\cite{greedan1971magnetic}, $^g$\cite{petit2018magnetic}}
\end{table}

\begin{figure}[!t]
\includegraphics[width=0.8\linewidth]{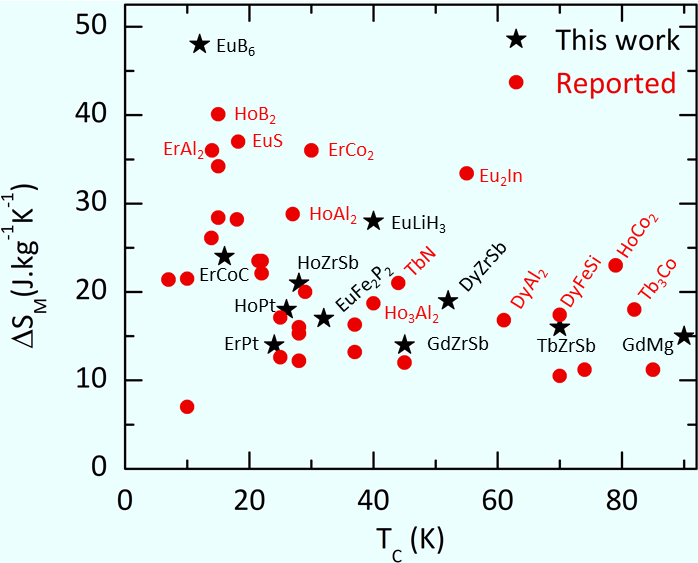}
\caption{\textbf{Comparison of newly discovered MCE materials to the known MCE materials.} The magnetocaloric entropy change ($\Delta$S$_M$) for the predicted compounds in this work (black stars) are plotted alongside some known MCE materials reported in the literature (red circles) in the 10\,K - 100\,K range for comparison.} 
\label{fig:fig-dSM}
\end{figure}

In the medium-$T_C$ range, EuLiH$_3$ exhibits a notable $\Delta S_M$ value of 28 J\,kg$^{-1}$K$^{-1}$ (0.15 J\,cm$^{-3}$K$^{-1}$) and a $\Delta T_{ad}$ of 6.7\,K at approximately 40\,K. These values are comparable to the values observed for TbN at 44 K ($\Delta S_M \approx$ 21 J\,kg$^{-1}$K$^{-1} \approx$ 0.20 J\,cm$^{-3}$K$^{-1}$, $\Delta T_{ad} \approx$ 8\,K in the $\langle 001 \rangle$ direction)~\cite{yamamoto2004magnetocaloric,von2013investigation}. Moving to the higher-$T_C$ range, DyZrSb displays the largest $\Delta S_M$ at 52\,K, while GdMg exhibits the smallest at 90\,K (refer to Table~\ref{tab:ASD}). Additionally, it is noteworthy that the predicted materials include a series with the general formula RZrSb (where R = Gd, Tb, Dy, and Ho), spanning a wide temperature range from 28\,K to 70\,K. Among these, HoZrSb stands out with the highest estimated $\Delta S_M$, attributed to the relatively larger effective moment associated with Ho compared to the other elements in the series. The predicted compounds are compared with other materials exhibiting significant magnetocaloric responses in the temperature range of 10\,K to 90\,K for a field change of 5\,T, as illustrated in Fig.~\ref{fig:fig-dSM}.

\section{Conclusion}
In this work, we exploited LLM's remarkable abilities to summarize texts and create an auto-generated MCE material database from public literature abstracts. Following curation and validation, we trained a neural network with high-quality data to predict good MCE materials. This model efficiently identified unexplored binary and ternary magnetic compounds in the Materials Project database, uncovering promising multiple materials across a wide range of temperatures. We then validated the top predictions using DFT simulations and ASD, discovering new MCE materials for the 10\,K to 100\,K temperature range. By integrating the results back into the database, we established a feedback loop that enhances the neural network's accuracy. This approach showcases the power of LLMs in mining literature for data-driven material discovery.

\section*{Data Availability Statement}

The data that supports the findings of this study are available within the article and its supplementary information. The auto-generated MCE database is available online at \url{https://github.com/RunqingYang1996/MCEdatabase}.

\begin{acknowledgments}
We thank Dr. Amir Jahromi and Dr. Ali Kashani for their helpful discussions. This work is based on research supported by the National Aeronautics and Space Administration (NASA) under award number 80NSSC21K1812. This work used Stampede2 at Texas Advanced Computing Center (TACC) through allocation MAT200011 from the Advanced Cyberinfrastructure Coordination Ecosystem: Services \& Support (ACCESS) program, which is supported by National Science Foundation grants 2138259, 2138286, 2138307, 2137603, and 2138296. Use was also made of computational facilities purchased with funds from the National Science Foundation (award number CNS-1725797) and administered by the Center for Scientific Computing (CSC) at the University of California, Santa Barbara (UCSB). The CSC is supported by the California NanoSystems Institute and the Materials Research Science and Engineering Center (MRSEC; NSF DMR-1720256) at UCSB. 
\end{acknowledgments}

\bibliography{references.bib}

\end{document}



\title{Supplementary Information: Enhancing Magnetocaloric Material Discovery: A Machine Learning Approach Using an Autogenerated Database by Large Language Models}

\author{Jiaoyue Yuan}
\thanks{These authors contributed equally.} \affiliation{Department of Mechanical Engineering, University of California, Santa Barbara, CA 93106, USA}
\affiliation{Department of Physics, University of California, Santa Barbara, CA 93106, USA}

\author{Runqing Yang}
\thanks{These authors contributed equally.} \affiliation{Department of Mechanical Engineering, University of California, Santa Barbara, CA 93106, USA}

\author{Lokanath Patra}
\thanks{These authors contributed equally.} \affiliation{Department of Mechanical Engineering, University of California, Santa Barbara, CA 93106, USA}

\author{Bolin Liao}
\email{bliao@ucsb.edu} \affiliation{Department of Mechanical Engineering, University of California, Santa Barbara, CA 93106, USA}

\date{\today}

\maketitle


\renewcommand{\thefigure}{S\arabic{figure}}
\renewcommand{\thetable}{S\arabic{table}}

\section{Machine Learning Model Generation}

\subsection{Training Data}

\begin{figure}[!htb]
\includegraphics[scale=0.45]{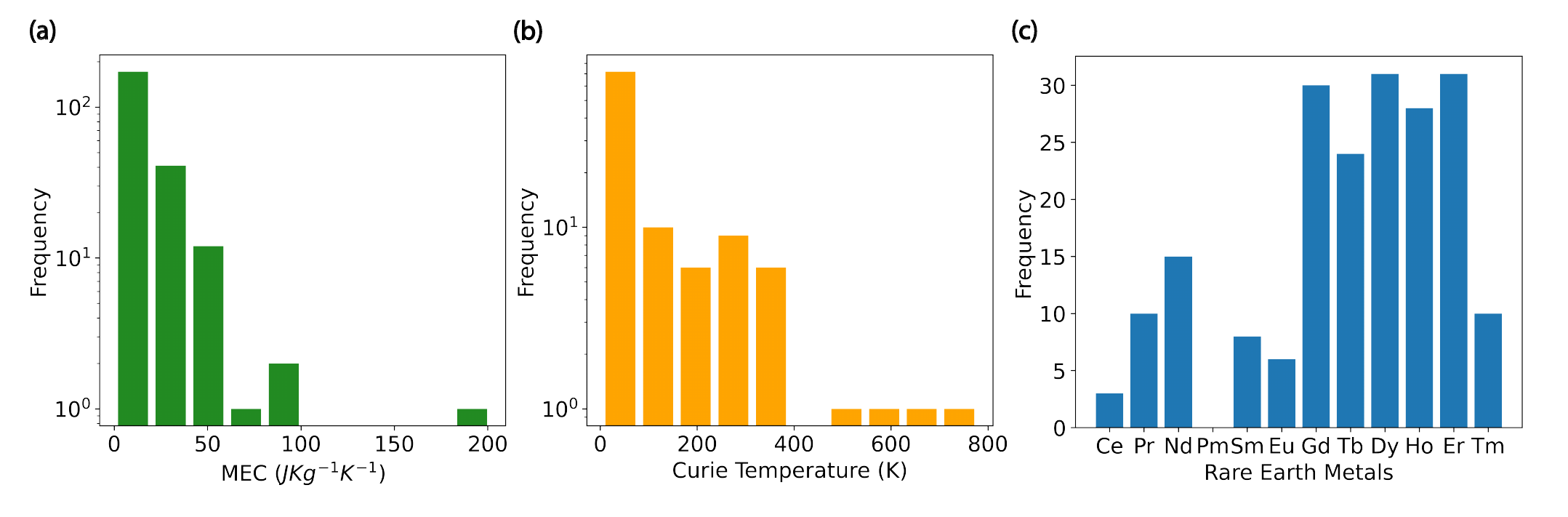}
\caption{Frequency distribution of (a) Magnetic Entropy Change, (b) Curie Temperature, and (c) rare-earth elements in the training data.} 
\label{fig:figS1}
\end{figure}

Figures~\ref{fig:figS1}(a) and (b) are the frequency distributions of $\Delta S_M$ and $T_C$ in the training data. Notably, a large portion of materials in the training dataset have a $T_C$ lower than 100\,K. This prevalent characteristic of the dataset is crucial for the model’s ability to discover more candidates with a $T_C$ in the desirable range. The frequency histogram of rare-earth elements shown in Fig.~\ref{fig:figS1}(c) reflects that the majority of reported MCE candidates contain heavy rare-earth metals.

\subsection{Model Training and Evaluation}
\begin{figure}[!tb]
\includegraphics[scale=0.65]{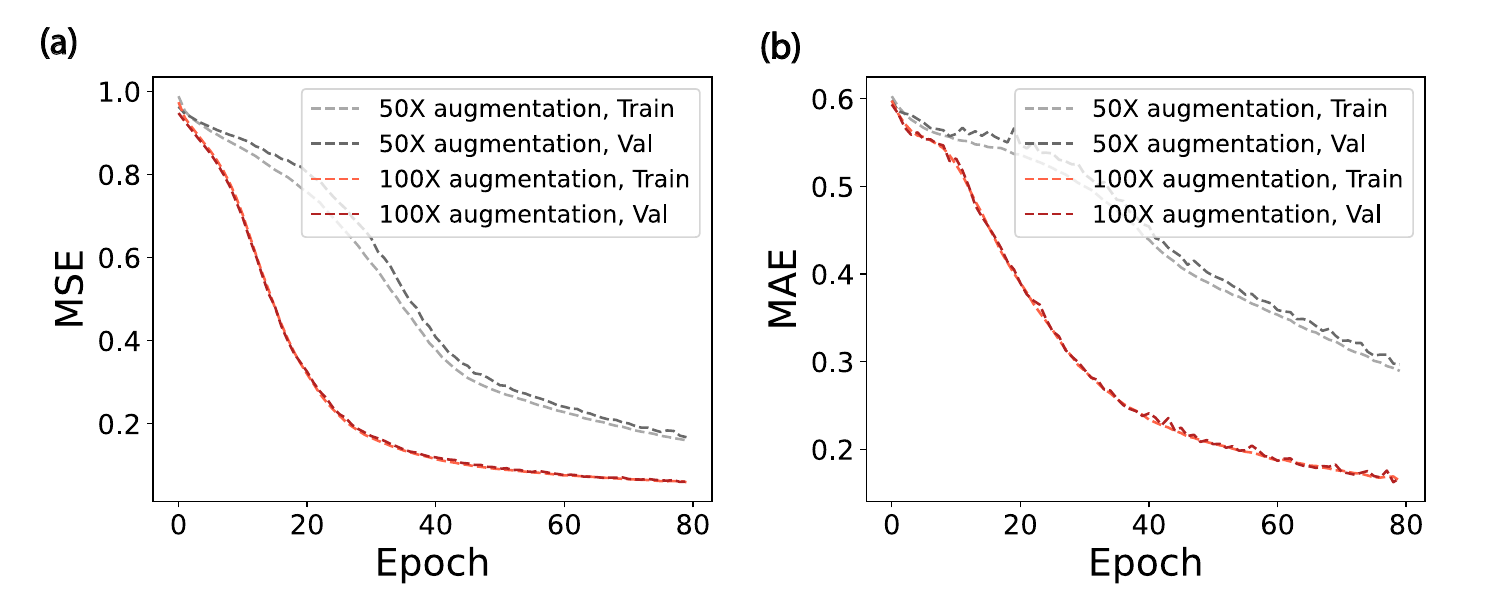}
\caption{Training curves of Mean-Squared Error (MSE) and Mean-Absolute Error (MAE) for augmentation factor of 50 vs. 100.} 
\label{fig:figS2}
\end{figure}

Figures~\ref{fig:figS2} (a) and (b) provide comparisons between the training curves for augmentation factors of 50 and 100, corresponding to datasets with 11,450 and 22,900 training data points, respectively. The model exhibits faster convergence when the training data size is doubled. This compelling evidence underscores the efficacy of the data augmentation method in facilitating the model's learning process. Intriguingly, despite the augmented data not introducing new information, the accelerated convergence suggests that the increased volume of training instances enhances the model's ability to generalize and capture intricate patterns within the dataset.

\begin{figure}[!htb]
\includegraphics[scale=0.5]{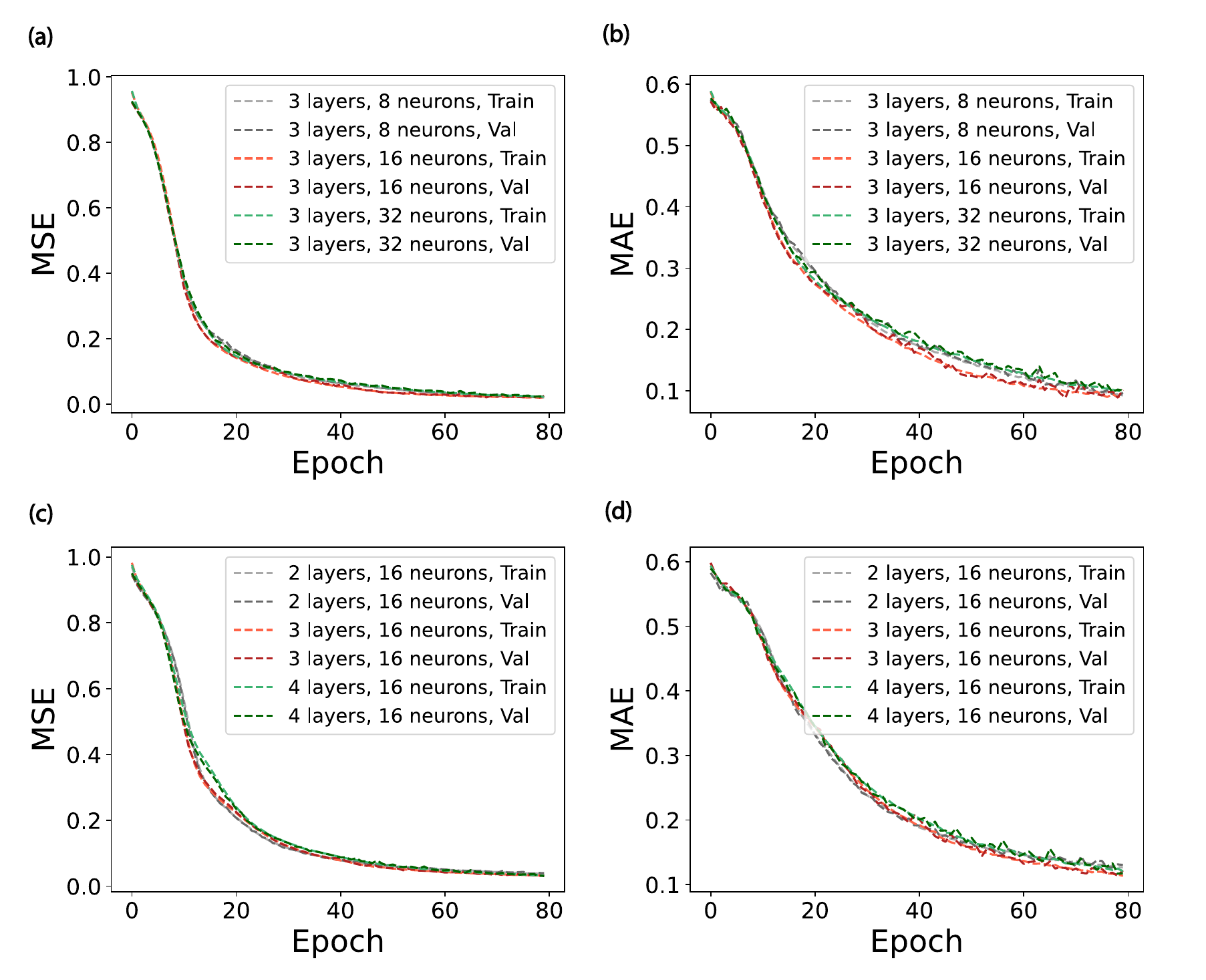}
\caption{Training curves of Mean-Squared Error (MSE) and Mean-Absolute Error (MAE) for grid search.} 
\label{fig:figS3}
\end{figure}

Figure~\ref{fig:figS3} displays a comprehensive analysis of the grid search results with fixed parameters of 3 layers and 16 neurons, respectively.  The training curves offer a visual representation of the model's performance across varying hyperparameter values of the network structure. For each training curve, the $x$-axis reflects the training epochs and the $y$-axis illustrates the corresponding values of Mean Squared Error (MSE) or Mean Absolute Error (MAE). It is noteworthy that MSE serves as the employed loss function. Across all configurations, MSE consistently demonstrates smooth convergence without discernible variations, indicating the stability of the model's learning process irrespective of hyperparameter adjustments. The nuanced analysis of MAE, however, provides valuable insights. Among the configurations tested, the networks consist of 3 layers with 16 neurons per layer emerging as the optimal structure.

